\documentclass[aps,prd,12pt,nofootinbib,superscriptaddress]{revtex4-2}
\usepackage{bm}
\usepackage{ulem}
\usepackage{comment}
\usepackage{amsmath}
\usepackage{amssymb}
\usepackage{booktabs}
\usepackage{graphicx}
\usepackage{float}
\usepackage{orcidlink}

\usepackage{hyperref}
\usepackage{CJK}
\usepackage{color}
\hypersetup{colorlinks=true, linkcolor=red, citecolor=blue}

\graphicspath{{figs/}}

\begin{document}
\begin{CJK*}{UTF8}{gbsn}

\title{
Dark matter distributions around extreme mass ratio inspirals: effects of radial pressure and relativistic treatment
}

\author{Yang Zhao \orcidlink{0009-0003-7436-8668}}
\email{zhaoyangedu@hust.edu.cn}
\affiliation{School of Physics, Huazhong University of Science and Technology, 1037 LuoYu Rd, Wuhan, Hubei 430074, China}

\author{Yungui Gong \orcidlink{0000-0001-5065-2259}}
\email{yggong@hust.edu.cn}
\affiliation{Institute of Fundamental Physics and Quantum Technology, Department of Physics, School of Physical Science and Technology,\\
Ningbo University, 818 Fenghua Rd, Ningbo, Zhejiang 315211, China}
\affiliation{School of Physics, Huazhong University of Science and Technology, 1037 LuoYu Rd, Wuhan, Hubei 430074, China}

\begin{abstract}
We investigate different treatments of dark matter (DM) distributions surrounding extreme mass ratio inspirals (EMRIs) to assess their impact on orbital evolution and gravitational wave emission. 
Density profiles derived from the mass current and from the energy-momentum tensor using a distribution function yield consistent results, 
but both differ substantially from profiles obtained using an anisotropic fluid model based on Einstein cluster ansatz. 
We find that the inclusion of radial pressure significantly modifies both the orbital dynamics and the resulting gravitational wave waveforms. 
By analyzing waveform dephasing and mismatches, we show that a fully relativistic treatment of DM distributions can substantially alter the detectability thresholds of DM halos. 
Our results demonstrate that radial pressure and relativistic modeling of DM are essential for accurately describing the dynamics and observational signatures of EMRIs embedded in DM halos.
\end{abstract}

\date{\today}
\maketitle
\end{CJK*}

\section{Introduction}
There is ample evidence for the existence of dark matter (DM), a key ingredient for understanding both the Universe and possible new physics \cite{Zwicky:1933gu, Rubin:1970zza, Rubin:1980zd, Bergstrom:1997fj, Allen:2002eu, Bertone:2004pz, Clowe:2006eq, Bertone:2016nfn, Bertone:2018krk, Planck:2018vyg}.
DM clusters around galaxies to form DM halos, so compact objects generally move in DM rich environments rather than in true vacuum \cite{Hernquist:1990be,Navarro:1994hi,Burkert:1995yz,Gondolo:1999ef}.
These halos can significantly affect the orbital evolution of compact binaries and thereby leave characteristic imprints on their emitted gravitational waves (GWs) \cite{Barausse:2014tra,Zwick:2025wkt}.
Conversely, precise GW measurements offer a unique probe of DM halo density profiles and microphysics \cite{Barack:2018yly,Coogan:2021uqv,Kavanagh:2020cfn,Cardoso:2021wlq,Figueiredo:2023gas,CanevaSantoro:2023aol}.

Since the first GW detection GW150914 in 2015 by the LIGO and Virgo collaborations  \cite{LIGOScientific:2016aoc,LIGOScientific:2016emj},
ground-based observatories have observed hundreds of stellar-mass black hole (BH) and neutron star mergers
\cite{LIGOScientific:2025hdt, LIGOScientific:2025slb,LIGOScientific:2025snk}.
Next-generation space-borne detectors, such as Laser Interferometer Space Antenna (LISA) \cite{Danzmann:1997hm, LISA:2017pwj,LISA:2024hlh}, Taiji \cite{Hu:2017mde}, TianQin \cite{TianQin:2015yph}, and the
Deci-hertz Interferometer GW Observatory (DECIGO) \cite{Kawamura:2011zz} will push sensitivity into the millihertz band
and enable unprecedented precision for probing black hole physics and fundamental physics \cite{Amaro-Seoane:2012vvq, Baibhav:2019rsa,LISA:2022kgy,LISA:2022yao,Gong:2021gvw,Zwick:2022dih}. 
In this band, extreme mass ratio inspirals (EMRIs, with mass ratio $\sim 10^{-4}-10^{-7}$), in which a stellar-mass compact object (SCO) inspirals into a massive black hole (MBH), are among the main sources \cite{Amaro-Seoane:2007osp,Babak:2017tow}.
During the last few years inspiralling deep in the strong gravitational field around the central MBH at highly relativistic speeds, 
the SCO completes roughly $10^6$ orbital cycles around the primary MBH \cite{Barack:2003fp}. 
This extensive mapping encodes the spacetime geometry of the primary with exquisite accuracy, offering a pristine opportunity to study astrophysical environments, cosmology, and gravity in the strong, nonlinear regime  \cite{Gair:2010yu,Yunes:2011ws, Kocsis:2011dr,Eda:2013gg,Berry:2019wgg, Destounis:2020kss,Maselli:2021men,Seoane:2021kkk,Laghi:2021pqk,Dai:2021olt,Cardoso:2022whc, Cole:2022yzw,Zhang:2024ugv,Dai:2023cft,Zhao:2024bpp}.
Consequently DM halos surrounding MBHs, 
which leave detectable imprints on the emitted GWs through their influence on orbital dynamics, 
can be probed with EMRIs, making accurate waveform modeling essential for EMRI science \cite{Barack:2009ux,Poisson:2011nh,Barack:2018yvs,Destounis:2021mqv,Warburton:2021kwk, Wardell:2021fyy, Isoyama:2021jjd,Rahman:2025mip,Zhang:2025eqz}.

In \cite{Cardoso:2021wlq,Cardoso:2022whc}, the authors generalized Einstein clusters to solve the Einstein equations sourced by Hernquist-type DM halos with a central MBH. 
The evolution of EMRIs in such halos has been studied for circular \cite{Figueiredo:2023gas, Zhang:2024ugv, Speeney:2024mas} and eccentric orbits  \cite{Cardoso:2020iji, Dai:2023cft}.
In the Einstein cluster ansatz, a strictly vanishing radial pressure $p_r=0$ is imposed \cite{Cardoso:2021wlq}.
Relaxing this assumption may be important for a fully self-consistent description.
On the other hand, the DM halo density is often modeled using the Hernquist profile \cite{Hernquist:1990be} at the Newtonian level in studies of DM halos around EMRIs \cite{Gondolo:1999ef} . 
A general relativistic analysis of DM halos around MBHs was presented in \cite{Sadeghian:2013laa}, 
and the results of \cite{Vicente:2025gsg} show that a fully relativistic treatment of DM halos  is required to model EMRIs in DM halos accurately. 
Using the BH perturbation method, GW fluxes for EMRIs embedded in DM halos were computed in \cite{Figueiredo:2023gas, Speeney:2024mas},
and first post-adiabatic (1PA) GW waveforms incorporating general relativistic treatment of DM halos have been developed \cite{Rahman:2025mip}.

In addition to GW radiation reaction, the motion of SCO  is also affected by gravitational drag (dynamical friction) \cite{Chandrasekhar:1943ys,Ostriker:1998fa,Kim:2007zb} and by accretion \cite{Bondi:1944rnk,Edgar:2004mk} from the ambient DM when the SCO moves through the DM medium. 
In this paper, we study circular EMRIs in galaxies enveloped by Hernquist-type DM halos using a fully relativistic treatment that retains both the halo density and a nonzero radial pressure. 
We evolve the SCO's orbit under the combined effects of GW radiation reaction and dynamical friction, 
and compare models with $p_r=0$, 
to quantify its impact on orbital evolution and the resulting GW waveform.

The paper is organized as follows. 
In Sec. II we review the energy density and pressure profiles of Hernquist-type DM halos using a fully relativistic treatment, 
and derive the background metric for EMRIs embedded in such halos. 
In Sec. III we solve the coupled evolution equations to obtain quasi-circular EMRI trajectories under the combined influence of dynamical friction and GW radiation reaction. 
In Sec. IV we generate GW waveforms with the Numerical Kludge (NK) method \cite{Babak:2006uv,Chua:2017ujo} and quantify the impact of the halo's radial pressure by computing mismatches between waveforms with and without that effect. 
We summarize our conclusions in Sec. V. 
Throughout this paper we use units $G = c = 1$.

\section{DM energy density and background geometry}
\subsection{DM energy density}
The phase space distribution function can be obtained from the Eddington inversion formula \cite{Eddington:1916}
\begin{equation}
\label{fE0}
f(E_0)=\frac{1}{\sqrt{8}\pi^2}\frac{d}{dE_0}\int_{E_0}^{0}d\Phi\frac{1}{\sqrt{\Phi-E_0}}\frac{d\rho_0}{d\Phi},
\end{equation}
where $\Phi$ is the Newtonian potential and $E_0$ is the Newtonian energy.
For the Hernquist-type DM halos, the density profile is \cite{Hernquist:1990be}
\begin{equation}
\label{HQ1}
\rho_0(r)=\frac{Ma_0}{2\pi r(r+a_0)^3},
\end{equation}
where $a_0$ is the typical halo size, and $M$ is the total halo mass.
The mass and the Newtonian potential of the halo are
\begin{equation}\label{MphiHQ-1}
\begin{split}
M(r)&=4\pi\int_{0}^{r}\rho_0(r)r^2dr= \frac{M r^2}{(a_0+r)^2},\\
\Phi(r)&=-\int_{r}^{\infty}dr\frac{M(r)}{r^2}=-\frac{M}{a_0+r}.
\end{split}
\end{equation}
Substituting Eqs. \eqref{HQ1} and \eqref{MphiHQ-1} into Eq. \eqref{fE0}, we get the phase space distribution function for the Hernquist model \cite{Lacroix:2018qqh}
\begin{equation}
\label{dmdfi}
\begin{split}
f_\text{HQ}(\epsilon_0)&=\frac{1}{\sqrt{2}(2\pi)^3(M a_0)^3}\frac{\sqrt{\epsilon_0}}{(1-\epsilon_0)^2}\\
&\times\left((1-2\epsilon_0)(8\epsilon_0^2-8\epsilon_0-3)+\frac{3\arcsin\sqrt{\epsilon_0}}{\sqrt{\epsilon_0(1-\epsilon_0)}}\right),
\end{split}
\end{equation}
where $\epsilon_0=-E_0 a_0/M$.

The mass current four-vector of DM halos around the central MBH is given by \cite{Sadeghian:2013laa}
\begin{equation}
J^\mu=\int d^4p \sqrt{-g}\frac{p^\mu}{m_h}f^{(4)}(p),
\end{equation}
where $m_h$ is the halo particle's rest mass, $p^\mu$ is the four momentum and $f^{(4)}(p)$ is the distribution function.
If all DM particles have the same mass $m_h$, then the distribution function can be written as 
\begin{equation}
\label{f4peq}
f^{(4)}(p)=\delta(m-m_h)f(\epsilon,L,L_z)m_h^{-3},
\end{equation}
where $\epsilon$, $L$, and $L_z$ represent the particle's energy, angular momentum and its $z$ component, respectively. 
From the relationship between the mass density and the mass current four-vector 
\begin{equation}
J^\mu = \rho_\text{J} u^\mu_\text{env},
\end{equation}
we get
\begin{equation}
\label{rhojeq1}
\rho_\text{J}=J^0/u^0_\text{env}=\sqrt{-g_{00}}\int d^4p \sqrt{-g}\frac{p^0}{m_h}f^{(4)}(p),
\end{equation}
where $\rho_\text{J}$ is the mass density as measured in a local freely falling frame, and $u^\mu_\text{env}=\sqrt{-g_{00}}(-1,0,0,0)$ is the four velocity of an element for DM halos.
For a spherically symmetric clusters, 
the energy $\epsilon$, angular momentum $L$ and its $z$ component $L_z$,
and the rest mass are conserved.
Using the conserved parameters ($\epsilon,\ L_z,\ L^2,\ m$), we get
\begin{equation}
d^4p=|J|^{-1}d\epsilon dL_z dL^2dm=\left|\frac{\partial(\epsilon,L_z,L^2,m)}{\partial(p_t,p_r,p_\theta,p_\phi)}\right|^{-1}d\epsilon dL_z dL^2dm.
\end{equation}
For a Schwarzschild BH of mass $M_\text{BH}$,
substituting Eq. \eqref{f4peq} into Eq. \eqref{rhojeq1}, 
we get the density of DM halos \cite{Sadeghian:2013laa}
\begin{equation}
\label{rhoHQ}
\begin{split}
\rho_\text{J}&=\frac{1}{\sqrt{2}(2\pi)^2}\frac{M}{a_0^3}\frac{a_0}{r-2M_\text{BH}}\int_{\bar{\epsilon}_\text{min}}^{\bar{\epsilon}_\text{max}}\left(1-\frac{M}{a_0}\bar{\epsilon}\right)d\bar{\epsilon}\\
&\times\int_{\bar{L}_\text{min}^2}^{\bar{L}_\text{max}^2}d\bar{L}^2\frac{\bar{f}(\bar{\epsilon})}{\sqrt{\bar{L}_\text{max}^2-\bar{L}^2}},
\end{split}
\end{equation}
where 
$\bar{\epsilon}=a_0(1-\epsilon)/M$, $\bar{L}=L/\sqrt{a M}$, $\bar{\epsilon}_{max}=a_0(1-\epsilon_\text{min})/M$, $\bar{\epsilon}_{min}=a_0(1-\epsilon_\text{max})/M$,
$\bar{L}_\text{max}=L_\text{max}/\sqrt{ a_0 M}$, $\bar{L}_\text{min}=L_\text{min}/\sqrt{ a_0 M}$
and $\bar{f}\left(\bar{\epsilon}\right)=\sqrt{2}(2\pi)^3(Ma_0)^{3/2} f(\bar{\epsilon})/M$.
The values of $L_\text{min}$, $L_\text{max}$, $\epsilon_\text{min}$ and $\epsilon_\text{max}$ are determined by the capture condition of the central MBH.
For the Schwarzschild BH of mass $M_\text{BH}$, $\epsilon_\text{max}=1$, so $\bar{\epsilon}_\text{min}=0$.
For $r>6 M_\text{BH}$, 
\begin{equation}
\bar{\epsilon}_\text{max}= \frac{a_0}{M} \left(1 - \frac{1 + 2M_\text{BH}/r}{(1 + 6M_\text{BH}/r)^{1/2}}\right).
\end{equation}
For $4M_\text{BH}<r<6 M_\text{BH}$, 
\begin{equation}
\bar{\epsilon}_\text{max}= \frac{a_0}{M} \left(1 - \frac{1 - 2M_\text{BH}/r}{(1 - 3M_\text{BH}/r)^{1/2}}\right).
\end{equation}
The results for $\bar{L}^2_\text{max}$ and $\bar{L}^2_\text{min}$ are
\begin{equation}
\begin{split}
\bar{L}^2_\text{max}&=\frac{r^2}{a_0M}\left(\frac{\epsilon^2}{1-2M_\text{BH}/r}-1\right),\\
\bar{L}^2_\text{min}&=\frac{32 M_\text{BH}^2/(a_0 M)}{36\epsilon^2-27\epsilon^4-8+\epsilon(9\epsilon^2-8)^{3/2}}.
\end{split}    
\end{equation}

Under adiabatic evolution, the action variables remain invariant,
\begin{equation}
\label{action}
\begin{split}
I_r^{fin}=I_r^{ini},\\
I_\theta^{fin}=I_\theta^{ini},\\
I_\phi^{fin}=I_\phi^{ini}.
\end{split}
\end{equation}
So we can obtain the relation between the initial and final stages for the parameters ($\epsilon,\ L_z,\ L^2$) from the invariance of action variables.
At the initial stage and in the Newtonian limit, the action variables are 
\begin{equation}
\label{Iinitial}
\begin{split}
I_r^{ini}&=\oint dr v_r=\oint dr\sqrt{2E-2\Phi-\frac{h}{r^2}},\\
I_\theta^{ini}&=\oint d\theta v_\theta=2\pi(h-h_z),\\
I_\phi^{ini}&=\oint d\phi v_\phi=2\pi h_z,
\end{split}
\end{equation}
where $v_r$, $v_\theta$ and $v_\phi$ are the velocity components of a DM particle, and $E$, $h$ and $h_z$ denote the energy, total angular momentum and its $z$ component at the initial stage, respectively.
In the final configuration, the DM halo is dominated by the potential of the central MBH of $M_\text{BH}$.
For a Schwarzschild spacetime, the action variables are \cite{Sadeghian:2013laa}
\begin{equation}\label{Ifinal}
\begin{split}
I_r^{fin}&=\int dr\sqrt{\epsilon^2-\left(1-\frac{2M_\text{BH}}{r}\right)\left(1+\frac{L^2}{r^2}\right)}\,,\\
I_\theta^{fin}&=2\pi(L-L_z),\\
I_\phi^{fin}&=2\pi L_z.
\end{split}
\end{equation}
From the invariance of action variables, we find that the angular momentum and its $z$ component are conserved, i.e., $h=L$, and $h_z=L_z$.
The energy at the initial stage can therefore be written as
$E=E(\epsilon,L)$ from the relation $I_r^{ini}(E,L)=I_r^{fin}(\epsilon, L)$.
Assuming that the distribution function is invariant under adiabatic evolution \cite{Quinlan:1994ed}, and taking the initial distribution function to be $f_\text{HQ}(E)$ from Eq. \eqref{dmdfi},
the final DM distribution function become
\begin{equation}
\label{ffin}
\bar{f}_\text{fin}(\bar{\epsilon})=\bar{f}_\text{HQ}\left(E(\epsilon,L)\right).
\end{equation}

Combining Eqs. \eqref{rhoHQ} and \eqref{ffin}, we get
\begin{equation}\label{rhoHQr}
\begin{split}
\rho_\text{J} = \frac{M\bar{\epsilon}_\text{max}}{\sqrt{2}(2\pi)^2a_0^2(r-2M_\text{BH})}\int_0^1\left(1-(M/a_0)\bar{\epsilon}_\text{max} u\right)du,\\
\times\int_0^1\sqrt{\frac{\bar{L}^2_\text{max}(u)-\bar{L}^2_\text{min(u)}}{1-z}}\,\bar{f}_\text{HQ}\left(E(\bar{\epsilon},\bar{L})\right)dz,
\end{split}
\end{equation}
where $\bar{L}^2=z\bar{L}^2_\text{max}+(1-z)\bar{L}^2_\text{min}$ and $u=\bar{\epsilon}/\bar{\epsilon}_\text{max}$.

We can also get the energy density $\rho_\text{T}$ and pressure for DM halos around a MBH from
the energy-momentum tensor,
\begin{equation}
\label{energymomentum1}
\begin{split}
T^{\mu\nu}&=\int f^{(4)}(p)p^\mu p^\nu \sqrt{-g}\,d^4 p\\
&=\rho_\text{T} u_\mu^\text{env} u_\nu^\text{env}+p_r k_\mu k_\nu+p_t(g_{\mu\nu}-k_\mu k_\nu+u_\mu^\text{env} u_\nu^\text{env}),
\end{split}
\end{equation}
from which
\begin{equation}\label{sigmapr}
\begin{split}
\rho_\text{T} &=T_{\mu\nu}u^\mu_\text{env} u^\nu_\text{env},\\
p_r&=T_{\mu\nu}k^\mu k^\nu,
\end{split}
\end{equation}
where $p_r$ is the radial pressure, $p_t$ is the tangential pressure,
and $k^\mu$ denotes a unit spacelike vector orthogonal to $u_\mu^\text{env}$, such that $k_\mu k^\mu=1$ and $u_\mu^\text{env} k^\mu=0$.
Combining Eqs. \eqref{energymomentum1}  and \eqref{sigmapr}, we get
\begin{equation}
\label{prHQ1}
\begin{split}
p_r(r)&=\frac{1}{\sqrt{2}(2\pi)^2\sqrt{1-2M/a_0(\bar{m}/x)}x^3}\frac{M^2}{a_0^4}\bar{\epsilon}_\text{max}\\
&\int_{0}^{1}\int_{0}^{1}\bar{f}_\text{HQ}\left(E(\bar{\epsilon},\bar{L})\right)\left(\bar{L}^2_\text{max}-\bar{L}^2_\text{min}\right)^{3/2}\sqrt{1-z}dzdu,
\end{split}
\end{equation}
\begin{equation}
\label{sigmaHQ1}
\begin{split}
\rho_\text{T} (r) &= \frac{1}{\sqrt{2}(2\pi)^2\,x\,\left(1-2M/a_0(\bar{m}/x)\right)^{3/2}}\frac{M}{a_0^3}\bar{\epsilon}_\text{max}\\
&\quad \int_0^1\int_0^1\sqrt{\frac{\bar{L}_\text{max}^2-\bar{L}_\text{min}^2}{1-z}}\bar{f}_\text{HQ}\left(E(\bar{\epsilon},\bar{L})\right)\left(1-u\bar{\epsilon}_\text{max}\frac{M}{a_0}\right)^2dudz,
\end{split}
\end{equation}
where $x=r/a_0$ and $\bar{m}=M_\text{BH}/M$.

On the other hand, assuming an anisotropic fluid with zero radial pressure for the DM halo, 
the matter density of a Hernquist-type halo surrounding a BH of mass $M_\text{BH}$ was found to be  \cite{Cardoso:2021wlq}
\begin{equation}
\label{cardosorho}
\rho_\text{C}=\frac{M (a_0+2M_\text{BH})}{2\pi r(r+a_0)^3}\left(1-\frac{2M_\text{BH}}{r}\right).
\end{equation}

\begin{figure}[htp]
    \centering{   \includegraphics[width=0.6\textwidth]{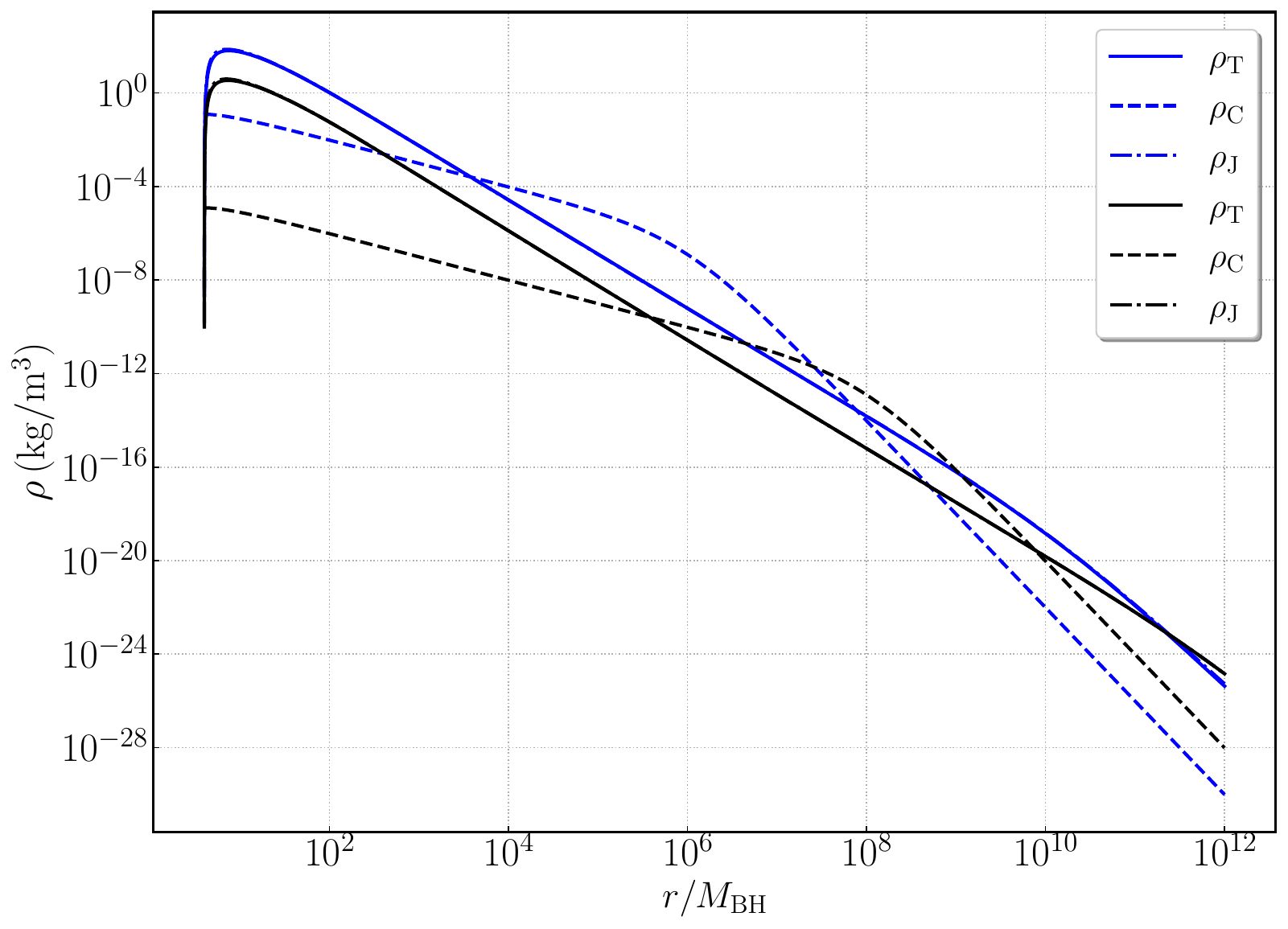}}
    \caption{
    The energy density distribution.
    The blue and black curves correspond to the halo size of $a_0 = 100 M$ and $a_0 = 10^4 M$, respectively. The solid, dashed and dot-dashed lines denote $\rho_\text{T}(r)$, $\rho_\text{C}(r)$ and $\rho_\text{J}(r)$, respectively.
    }
\label{fig:1}
\end{figure}

\begin{figure}[htp]
    \centering{   \includegraphics[width=0.9\textwidth]{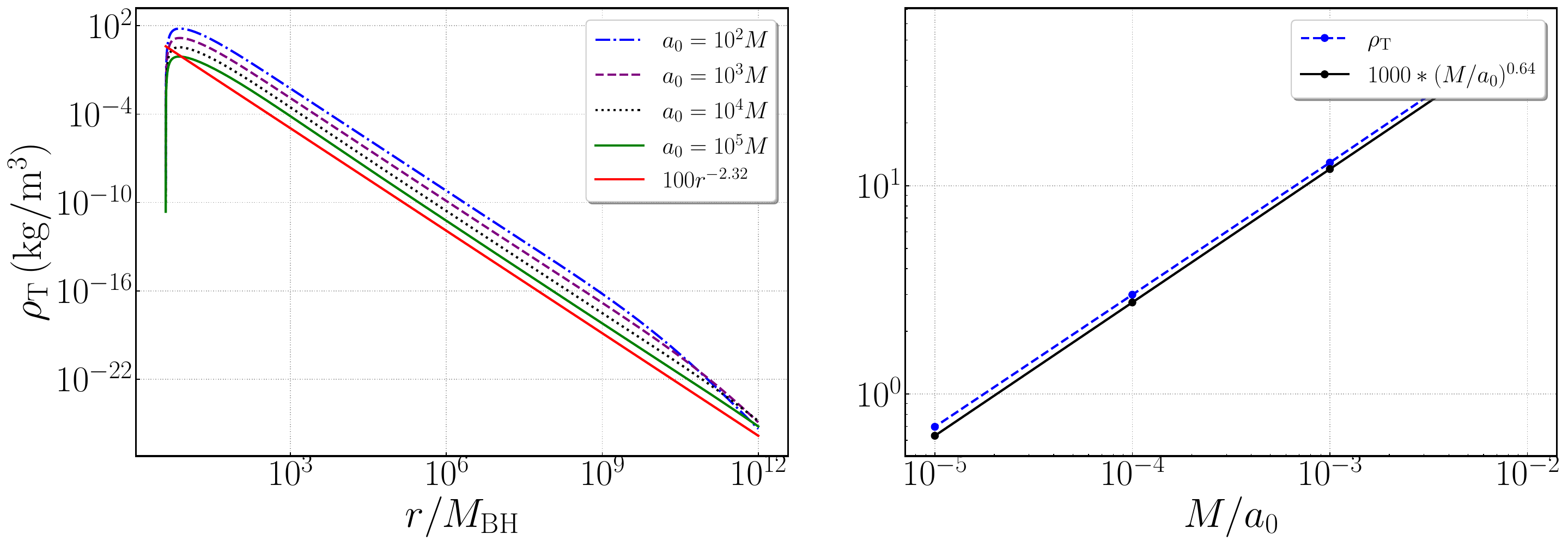}}
    \caption{
    The dependence of $\rho_\text{T}$ on $r$ and the compactness $M/a_0$. The right panel is plotted at $r=10M_\text{BH}$.
    }
\label{fig:1b}
\end{figure}

\begin{figure}[htp]
    \centering{
    \includegraphics[width=0.6\textwidth]{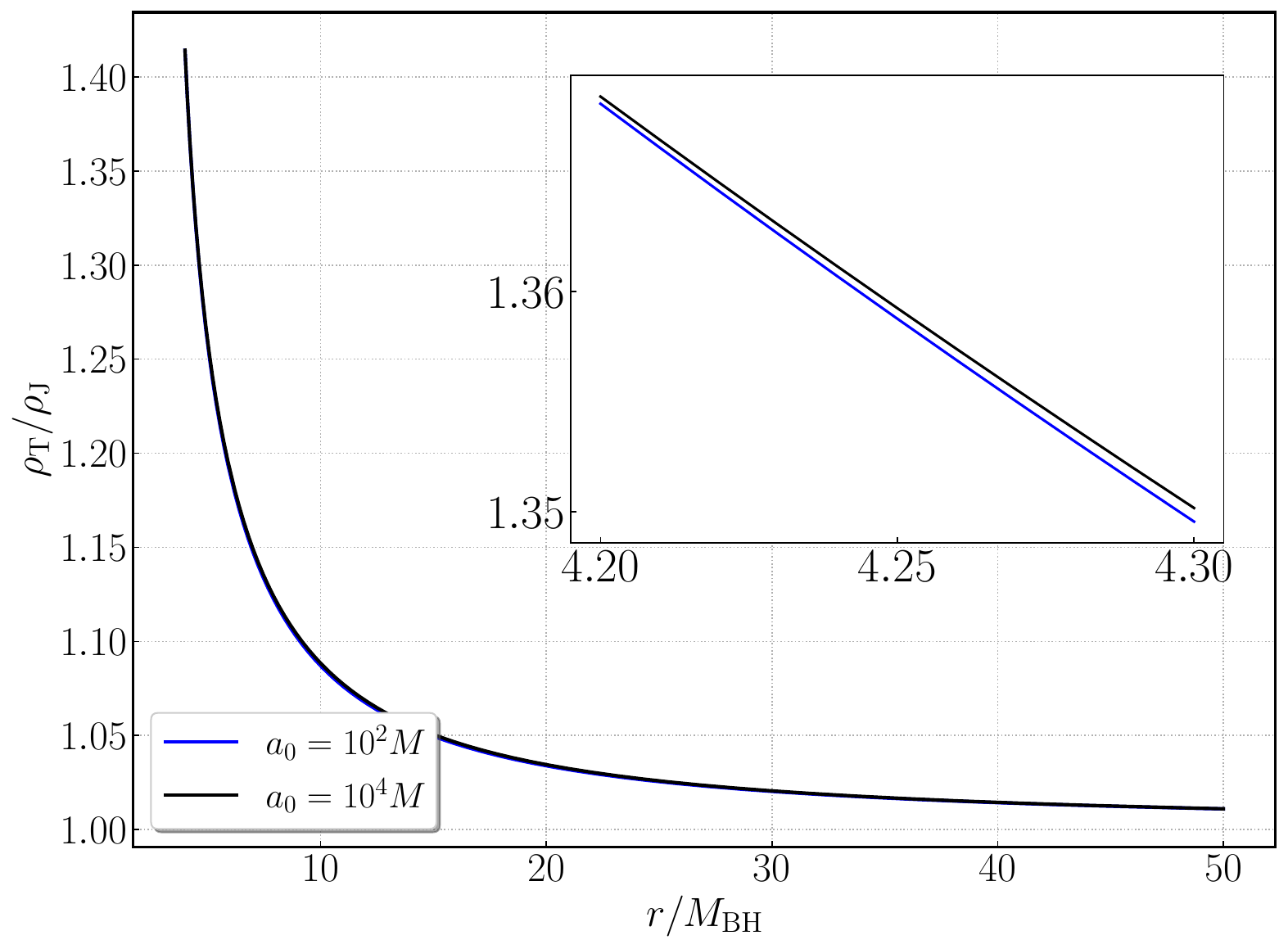}}
    \caption{
    The density ratio $\rho_\text{T}/\rho_\text{J}$ near the central MBH.
    The blue and black curves correspond to the halo size of $a_0 = 100 M$ and $a_0 = 10^4 M$, respectively.
    }
\label{fig:1a}
\end{figure}

Without loss of generality, we choose $M_\text{BH}=10^6 M_\odot$ and the initial mass of the Hernquist-type DM halo $M=10^4 M_\text{BH}$ in this paper.
Take $a_0=100M$ or $a_0=10^4 M$,
we plot the energy densities
$\rho_\text{J}(r)$, $\rho_\text{T}(r)$ and $\rho_\text{C}(r)$ in Fig. \ref{fig:1}.
From Fig. \ref{fig:1}, we see that $\rho_\text{J}$ closely follows $\rho_\text{T}$,
while both differ significantly from $\rho_\text{C}$.
For the compactness values $M/a_0=10^{-4}$ and $M/a_0=10^{-2}$ , 
the peak value of $\rho_\text{T}$  exceeds that of $\rho_\text{C}$
by about six orders of magnitude and three orders of magnitude, respectively, at the same radius. 
Far from the central MBH the ratio $\rho_\text{C}/\rho_\text{T}$  (and $\rho_\text{C}/\rho_\text{J}$) is much less than 1; it rises above 1 at intermediate radii, but drops to much smaller than 1 again near the MBH. 
As shown in Fig. \ref{fig:1b}, $\rho_\text{T}\sim r^{-2.32}$ and $\rho_\text{T}\sim (M/a_0)^{0.64}$ around the peak. 
Consequently, $\rho_\text{T}$ falls off faster than $\rho_\text{C}$ with increasing $r$ and its peak value is much larger than $\rho_\text{C}\sim M/a_0^2$.
To highlight the difference between $\rho_{\rm J}$ and $\rho_{\rm T}$, 
we plot their relative difference near the peak in Fig.\ \ref{fig:1a}. We find that $\rho_\text{T}/\rho_\text{J}$ increases as $r$ decreases, reaching $1.4$ near the horizon.

\begin{figure}[htp]
    \centering{
    \includegraphics[width=0.6\textwidth]{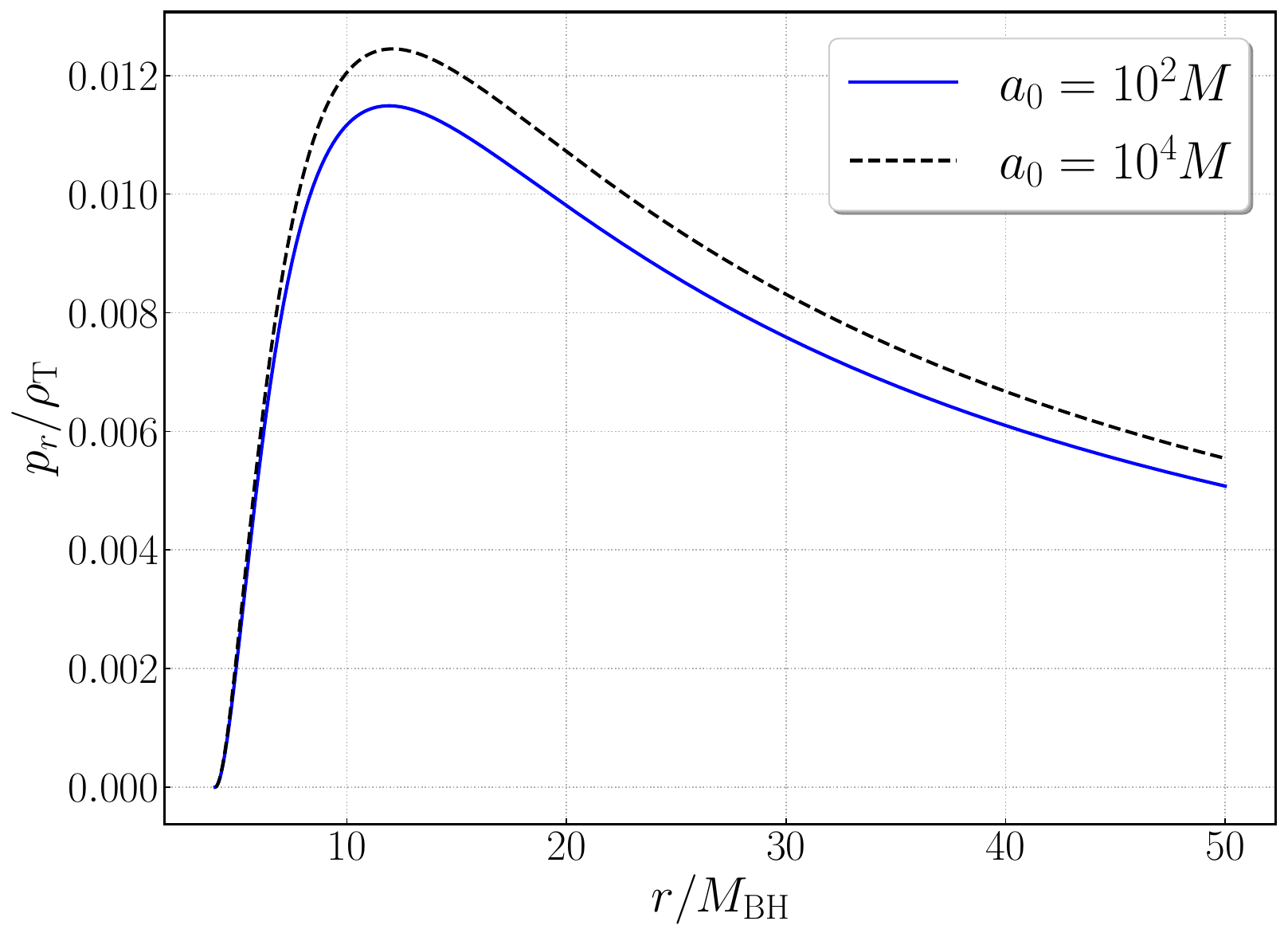}}
    \caption{
    The ratio between $p_r$ and $\rho_\text{T}$. 
    The blue and black curves correspond to the halo size of $a_0 = 100 M$ and $a_0 = 10^4 M$, respectively.
    }
\label{fig:2}
\end{figure}

The ratio $p_r/\rho_\text{T}$ is shown in Fig. \ref{fig:2}. 
From the figure we observe that the maximum ratio is $\approx 0.01$ for both $a_0=100M$  and  $a_0=10^4 M$. 
As $r$ decreases the ratio rises to this maximum near $r\approx 10 M_\text{BH}$  and then falls off sharply closer to the central MBH.

\subsection{Background metric}
After deriving the DM distribution, we now solve Einstein's equation to obtain the background geometry.
For the energy-momentum tensor \eqref{energymomentum1}, we consider three cases: (1) $\rho=\rho_\text{T}$ and $p_r$ calculated from Eqs. \eqref{prHQ1} and \eqref{sigmaHQ1}; (2) $\rho=\rho_\text{J}$ calculated from Eq. \eqref{rhoHQr} and $p_r=0$; (3) $\rho=\rho_\text{C}$ calculated from Eq. \eqref{cardosorho} and $p_r=0$.
The key difference between cases 1 and 2 is the presence of radial pressure in the former.
Taking a static and spherically symmetric metric 
\begin{equation}
\label{metric}
ds^2=-e^{2A(r)}dt^2+\frac{dr^2}{1-2m(r)/r}+r^2d\Omega^2,
\end{equation}
and solving Einstein's equation with the energy-momentum tensor \eqref{energymomentum1}, we get
\begin{equation}
\label{Einstein equation}
\begin{split}
A'(r)&=\frac{m(r)+4\pi r^3p_r(r)}{r(r-2m(r))},\\
m'(r)&=4\pi r^2\rho(r),
\end{split}
\end{equation}
where $m(r)$ is the mass function and a prime indicates differentiation with respect to $r$.

The metric can be solved by using the following boundary conditions
\begin{equation}
\label{boundary}
\begin{split}
m(R_s)&=M_\text{BH},\\
A(r\rightarrow r_\text{out})&=\ln\left(1-\frac{2m(r)}{r}\right),
\end{split}
\end{equation}
where $R_s=2M_\text{BH}$ is the event horizon of the central MBH, and $r_\text{out}$ corresponds to spatial infinity.
For the numerical approximation of infinity we set
$r_\text{out}=10^6 a_0$ \cite{Figueiredo:2023gas, Konoplya:2022hbl}
and numerically integrate  Eqs. \eqref{Einstein equation} and \eqref{boundary} to obtain the background metric.

\section{The evolution of EMRIs}
For a SCO inspiralling into a MBH surrounded by DM halos, the energy and angular momentum of the SCO are
\begin{equation}\label{EandL}
\begin{split}
E_p=\mu\epsilon_p=\mu \sqrt{\frac{e^{2A(r)}}{1-r A'(r)}},\\
L_p=\mu h_p=\mu \frac{r\sqrt{rA'(r)}}{\sqrt{1-r A'(r)}},
\end{split}
\end{equation}
where $\mu$ is the mass of the SCO, 
and the subscript $p$ denotes the SCO.
We take $\mu=10M_\odot$ in this paper. 
Using the metric \eqref{metric}, we get the orbital frequency
\begin{equation}
\omega=e^{A(r)}\sqrt{\frac{A'(r)}{r}},
\end{equation}
where $v=\omega r$ is the velocity of the SCO.

When the small SCO moves through the meduim, it experiences a drag force known as dynamical friction \cite{Chandrasekhar:1943ys}.
For a BH in a circular orbit, the dynamical friction force is \cite{Ostriker:1998fa,Kim:2007zb,Barausse:2007ph}
\begin{equation}
f_\text{DF}^i=-\frac{4\pi\mu \ln\Lambda \rho(r) v^i}{v^3}\xi(v),
\end{equation}
where the subscript ``DF" denotes dynamical friction. 
The Coulomb logarithm is taken as $\ln\Lambda=3$ \cite{Eda:2014kra}. The factor \(\xi(v)=(1-\zeta v^2)(1+\zeta v^2)^2\) encodes the relativistic correction to the dynamical friction \cite{Speeney:2022ryg}. 
This correction is necessary for two reasons: (1) relativistic effects increase the gravitational deflection angle of DM particles encountered by the SCO compared with the classical prediction, and (2) the relativistic momentum of the DM particles must be accounted for. 
The parameter $\zeta$ switches this correction on or off: $\zeta=1$ includes the full relativistic effects, while $\zeta=0$ recovers the classical (nonrelativistic) expression.
The energy-loss rate due to dynamical friction is given by
\begin{equation}
\label{dEdtDF}
\left(\frac{dE}{dt}\right)_\text{DF}=\dot{E}_\text{DF}=f_\text{DF}^i v_i\,.
\end{equation}

\begin{figure}[htp]   \centering{\includegraphics[width=0.6\textwidth]{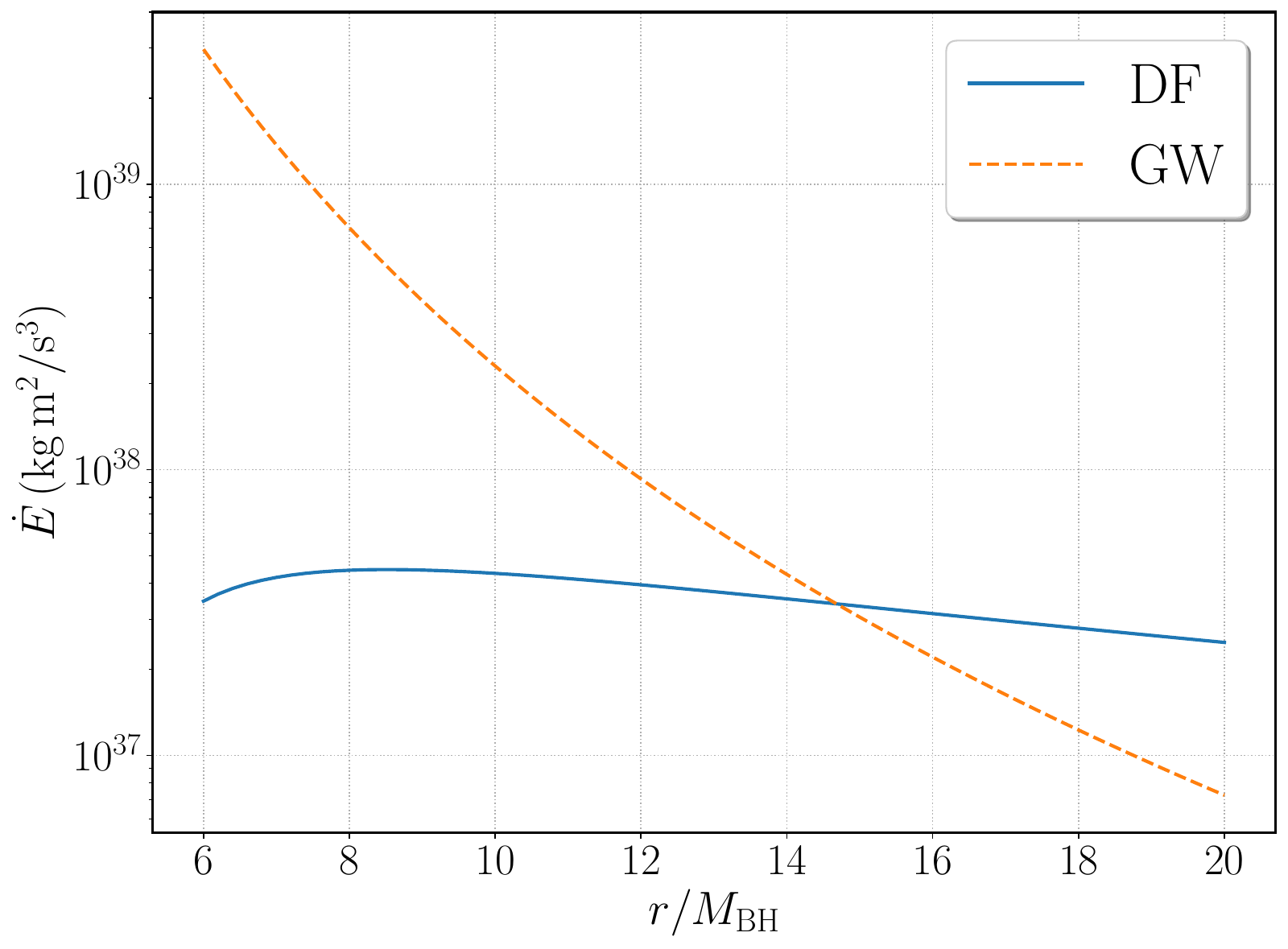}
    }
    \caption{
    The energy fluxes due to the dynamical friction and gravitational radiation from GWs. We take $a_0=100M$.
    }
\label{fig:4}
\end{figure}

The energy loss rate due to gravitational radiation from GWs, in the quadrupole approximation, is 
\begin{equation}
\label{dEdtGW}
\left(\frac{dE}{dt}\right)_\text{GW}=\dot{E}_\text{GW}=-\frac{32 \mu^2r^4\omega^6}{5}.
\end{equation}
Eqs. \eqref{dEdtDF} and \eqref{dEdtGW} allow a direct comparison of the energy-loss rates due to dynamical friction and gravitational radiation, 
and the results for the case 1 ($\rho_\text{T}$) with $a_0=100M$ are shown in Fig. \ref{fig:4}.
From the figure, the energy flux due to dynamical friction shows little change.  When $r<14M_\text{BH}$, 
the GW energy flux exceeds the dynamical friction flux, whereas for $r>16M_\text{BH}$, 
the GW flux is smaller than the dynamical friction flux,
so GW radiation dominates the energy loss near the central MBH. 
Results for other values of the compactness $M/a_0$ are similar.

\begin{figure}[htp]   \centering{\includegraphics[width=0.6\textwidth]{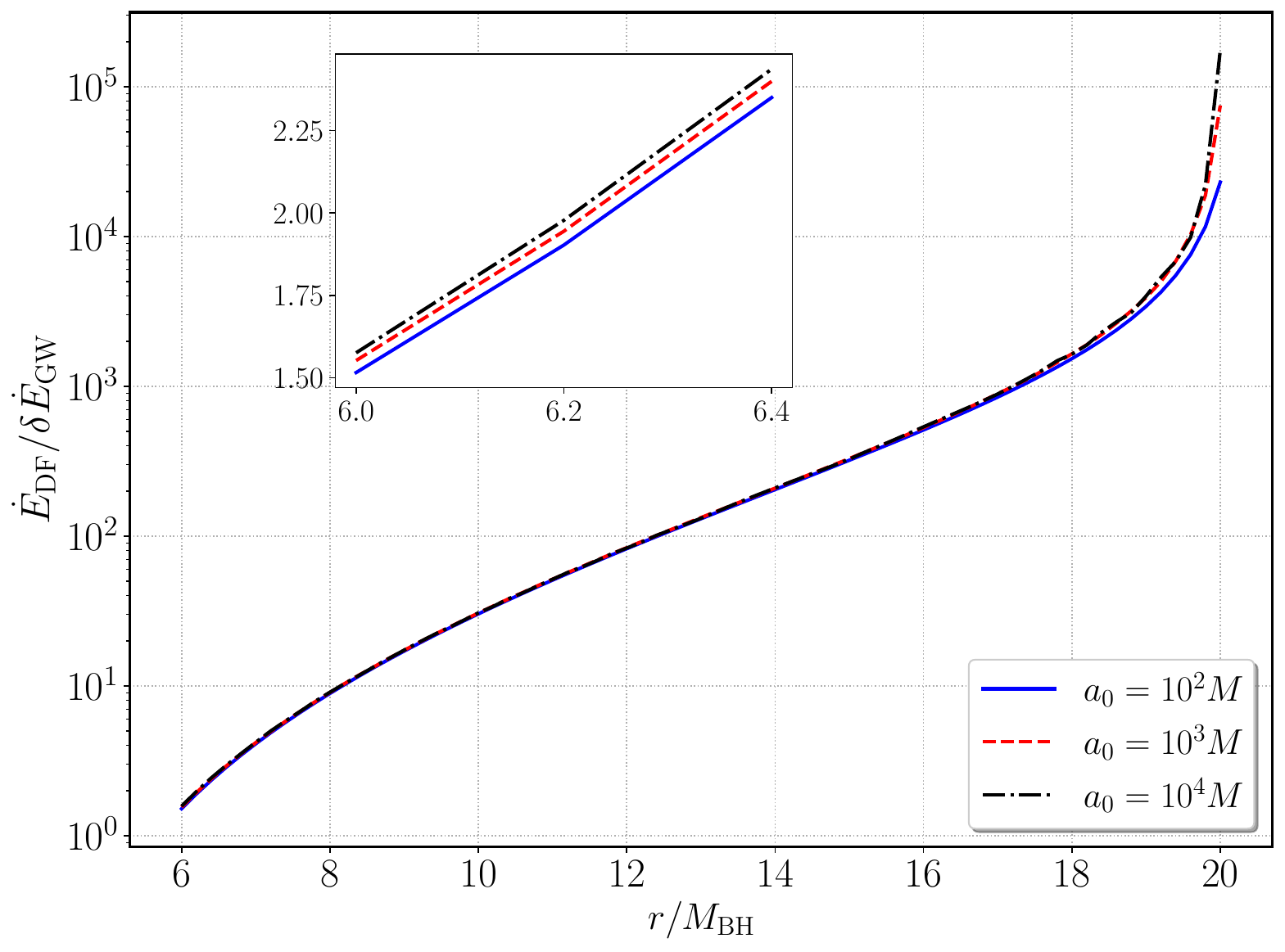}}
    \caption{
    Evolution of the ratio $\dot{E}_\text{DF}/\delta\dot{E}_\text{GW}$ as a function of radius $r$. The solid, dashed and dot-dashed lines correspond to $a_0=100M$, $a_0=1000M$ and $a_0=10^4M$, respectively.
    }
\label{fig:5}
\end{figure}

To separate the halo-induced contribution to the GW radiation, we write
\begin{equation}
\dot{E}_\text{GW}=\dot{E}_\text{GW0}+\delta\dot{E}_\text{GW},
\end{equation}
where $\dot{E}_\text{GW0}$ denotes the GW radiation of the EMRI in vacuum, and $\delta\dot{E}_\text{GW}$ the additional energy loss induced by the DM halo relative to the vacuum case. 
Figure  \ref{fig:5}  compares  $\delta\dot{E}_\text{GW}$ and $\dot{E}_\text{DF}$ for case 1 ($\rho_\text{T}$).
The figure shows that the dynamical friction loss $\dot{E}_\text{DF}$ is always larger than the halo-induced correction $\delta\dot{E}_\text{GW}$.
For $r<10M_\text{BH}$, the ratio $\dot{E}_\text{DF}/\delta \dot{E}_\text{GW}$ is nearly the same for different halo compactness values, though it is slightly smaller for larger compactness.

\begin{figure}[htp]
    \centering{
    \includegraphics[width=0.9\textwidth]{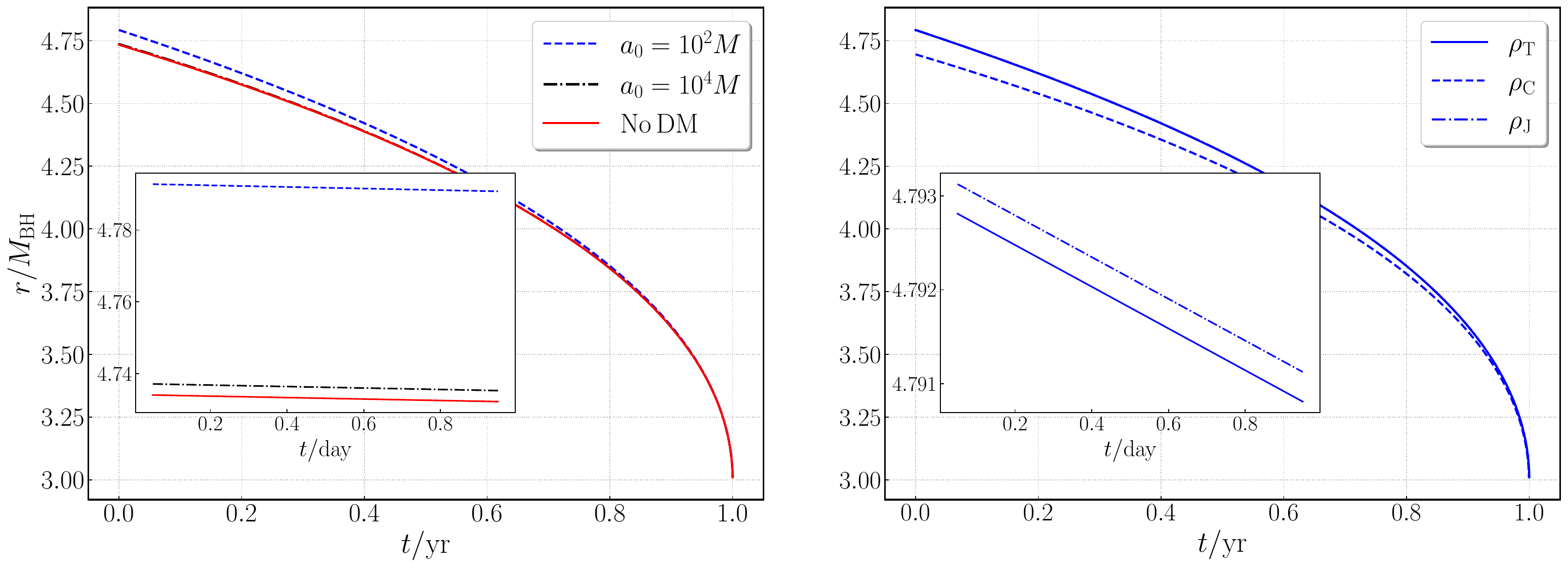}
    }
    \caption{Evolution of $r$  during the final year before the innermost stable circular orbit. In the left panel, the solid line shows the EMRI without a DM halo; the dashed line corresponds to a  DM halo for the case 1 ($\rho_\text{T}$) with $a_0=100M$; and the dot-dashed line corresponds to the case 1 ($\rho_\text{T}$) with $a_0=10000M$.
    The halo size is chosen as $a_0=100M$ in the right panel.
    }
\label{fig:6}
\end{figure}

Neglecting accretion from the surrounding DM halo, the total energy loss is 
\begin{equation}
\label{dEdtorb}
\left(\frac{dE}{dt}\right)_\text{orb}=\left(\frac{dE}{dt}\right)_\text{DF}+\left(\frac{dE}{dt}\right)_\text{GW}.
\end{equation}
The evolution of the orbital radius $r$ is therefore
\begin{equation}
\label{orb}
\dot{r}=\left(\frac{\dot{E}}{E'(r)}\right)_\text{orb}.
\end{equation}
We evolve the SCO orbit numerically and the results are shown in Fig. \ref{fig:6}. 
It is apparent that the presence of the DM halo accelerates the inspiral, with larger compactness $M/a_0$ yielding faster evolution. 
This acceleration is due to the additional energy loss from dynamical friction and the halo-induced enhancement of GW emission as shown in Fig. \ref{fig:5}.
Together, they cause a more rapid decrease in orbital energy, leading to a faster orbital evolution.
Figure \ref{fig:6} also shows that the orbital decay is fastest for the case 2 with $\rho_\text{J}$, followed by the case 1 with $\rho_\text{T}$, and slowest for the case 3 with $\rho_\text{C}$.
Comparing cases 1 and 2 indicates that radial pressure slightly slows the orbital decay.

\section{GW Waveforms}
As discussed above, DM halos alter the orbital dynamics of EMRIs, leaving measurable imprints on their GW waveforms.
The quadrupole formula for GWs is given by
\begin{equation}
h^{jk}=\frac{2}{d_\text{L}}\ddot{I}^{jk},
\end{equation}
where $d_L$ is the luminoslty distance from the detector to source and $I^{jk}=\mu x^j x^k$ is EMRI's quadrupole.
The plus and cross polarization modes of GWs are
\begin{equation}
    h_+=\frac{1}{2}(e_\text{X}^je_\text{X}^k-e_\text{Y}^je_\text{Y}^k)h_{jk},
\end{equation}
\begin{equation}
    h_\times=\frac{1}{2}(e_\text{X}^je_\text{Y}^k+e_\text{Y}^je_\text{X}^k)h_\text{jk},
\end{equation}
where $e_\text{X}$ and $e_\text{Y}$ are orthonormal vectors in the plane perpendicular to the direction from the detector to the GW source.
For circular orbits, GW polarizations are \cite{Wagoner:1976am}
\begin{equation}
\begin{split}
h_+&=\frac{4 r^2}{d_\text{L}}\frac{1+\cos\iota^2}{2}\mu\,\omega^2\cos(2\varphi),\\
h_\times&=-\frac{4 r^2}{d_\text{L}}\cos\iota\,\mu\,\omega^2\sin(2\varphi),
\end{split}
\end{equation}
where $\iota$ is the inclination angle between the binary's orbital angular momentum and the line of sight,
the orbital phase
\begin{equation}
\varphi=\varphi_0+\int_{t_{ini}}^{t_{fin}}\omega(t)dt,
\end{equation}
and $\varphi_0$ is the initial orbital phase.

\begin{figure}[htp]
    \centering{
    \includegraphics[width=0.6\textwidth]{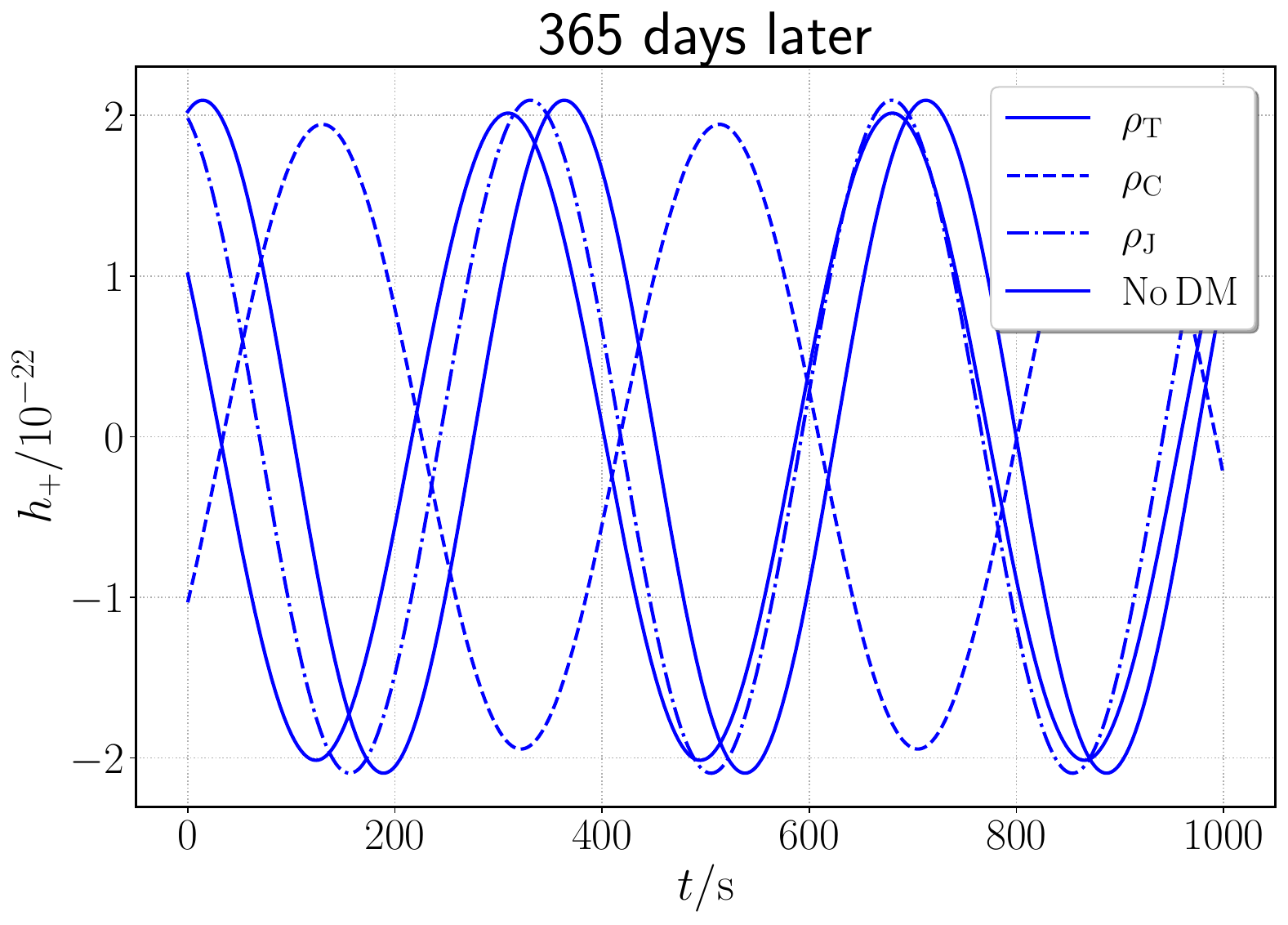}}
    \caption{
    The GW waveforms for EMRIs in the presence and absence of DM halo after  one year of evolution.
    }
\label{fig:7}
\end{figure}

Using the orbital evolution results from the previous section
and  taking the initial semi-latus rectum $p_0=10M_\text{BH}$,
the inclination angle $\iota=\pi/6$, 
the luminosity distance $d_\text{L}=1$ Gpc, and the initial longitude of pericenter $\omega_0=0$,
we compute the GW waveform for $a_0=100M$, and the plus polarization is shown in Fig. \ref{fig:7}. 
After one year of evolution as shown in Fig. \ref{fig:7}, 
the EMRI waveforms in DM halos are clearly distinguishable from those in vacuum.

To quantify the impact of DM halos on EMRI evolution, 
we compute the accumulated number of orbital cycles for EMRIs evolving with and without DM. 
The difference in number of cycles $\mathcal{N}$ is defined as $\Delta\mathcal{N}=\mathcal{N}_\text{DM}-\mathcal{N}_\text{vac}$. 
Following \citep{Maselli:2020zgv}, we adopt $\Delta\mathcal{N}\sim1\ \text{rad}$ accumulated over one year as the threshold for detectable dephasing.

Table \ref{table:1} lists $\Delta\mathcal{N}$ for (i) case 1 ($\rho_\text{T}$) versus vacuum, (ii) case 2 ($\rho_\text{J}$) versus vacuum, and (iii) case 3 ($\rho_\text{C}$) versus vacuum. 
For $M_\text{BH}=10^6 M_\odot$, $\mu=10M_\odot$ and  $M=10^4 M_\text{BH}$, DM halos with compactness $M/a_0>10^{-4}$ are detectable for case 3, 
whereas for cases 1 and 2 DM halos are detectable already for $M/a_0>10^{-5}$. 
Thus fully relativistic treatment on DM distribtuion ( $\rho_\text{T}$ or $\rho_\text{J}$  versus $\rho_\text{C}$) significantly affects the detectability.

Table \ref{table:1} also summarizes the differences among cases 1, 2, and 3. 
At $M/a_0=10^{-4}$, 
the cycle difference between cases 1 and 2 is $\delta\mathcal{N}\sim1\ \text{rad}$, indicating that radial pressure must be included.
The comparisons between cases 1 and 3 and between cases 2 and 3 further imply that a fully relativistic treatment of the DM distribution is necessary when $M/a_0<10^{-5}$.

\begin{table}[htp]
\centering
\caption{
Differences in accumulated orbital cycles for EMRIs in DM halos over one year. 
Left three columns show each case compared with the vacuum case; right three columns show pairwise differences among the three cases. }
\begin{tabular}{|c|c|c|c||c|c|c|}
\hline
$M/a_0$ & case 1 & case 2 &case 3 & case 1/2 & case 1/3 & case 2/3\\
\hline
$10^{-2}$ & $2604.0$ & $2627.6$ & $213.7$ & $23.6$ & $2390.4$ & $2414.0$ \\
$10^{-3}$ & $593.7$ & $599.4$ & $26.9$ & $5.7$ & $566.8$ & $572.5$ \\
$10^{-4}$ & $139.0$ & $139.9$ & $2.8$ & $0.9$ & $136.1$ & $137.0$ \\
$10^{-5}$ & $31.8$ & $32.9$ & $0.4$ &  & $31.4$ & $32.4$\\
\hline          
\end{tabular}
\label{table:1}
\end{table}

The orbital-cycle threshold alone maybe insufficient for assessing detectability, therefore we compute the mismatch between GW signals to discriminate more precisely among different GW waveforms.
The mismatch between two signals is
\begin{equation}
\label{Mismatch}
\text{Mismatch}[h_1, h_2] = 1-\text{Max}_{(t_0,\phi_0)}\frac{\left< h_1|h_2 \right>}{\sqrt{\left< h_1\vert h_1 \right> \left< h_2\vert h_2 \right>}},
\end{equation}
where the $(t_0,\phi_0)$ are the time and phase offsets \cite{Lindblom:2008cm},  
the inner product between two waveforms $h_1$ and $h_2$ are
\begin{equation}
    \langle h_1\vert h_2 \rangle=2\int_{f_\text{min}}^{f_\text{max}}\frac{\tilde{h}_1(f)\tilde{h}_2^*(f)+\tilde{h}_2(f)\tilde{h}_1^*(f)}{S_n(f)}df,
\end{equation}
$\tilde{h}(f)$ is the Fourier transformation of the time-domain signal $h(t)$, $\tilde{h}^*(f)$ is its complex conjugate,
$f_\text{min}$ and $f_\text{max}$ are
\begin{equation}
\begin{split}
    f_\text{min}&=\text{Min}(f_\text{end},f_\text{up}),\\
    f_\text{max}&=\text {Max}(f_\text{ini},f_\text{low}),
\end{split}
\end{equation}
$f_\text{ini}$ and $f_\text{end}$ are the initial and final frequencies for the orbital evolution, 
the lower and upper cutoff frequencies for LISA are chosen as $f_\text{low}=10^{-4}$  Hz and $f_\text{up}=1$ Hz, respectively  \cite{Yagi:2009zm},
and $S_n(f)$ is the noise spectral density for GW detectors.
The one-side noise power spectral density of space-borne GW detector is \cite{Robson:2018ifk}
\begin{equation}\label{psd-lisa}
\begin{split}
    S_n (f) =&\frac{S_x}{L^2}+\frac{2S_a \left[1+\cos^2(2\pi f L/c)\right]}{(2\pi f)^4 L^2}\\
    &\quad \times\left[1+\left(\frac{4\times 10^{-4}\text{Hz}}{f}\right) \right].
\end{split}
\end{equation}
For LISA, the arm length is $L=2.5\times 10^9$ m, 
the displacement noise is $\sqrt{S_x}=1.5\times 10^{-11}\text{ m}\,\text{Hz}^{-1/2}$ and the acceleration noise is $\sqrt{S_a}=3\times 10^{-15} \text{ m}\,\text{s}^{-2}\,\text{Hz}^{-1/2}$.

The signal-to-noise ratio (SNR) is $\text{SNR} = \left< h\vert h \right>$.
For the GW source  parameters chosen in this paper we get SNR$=23.8$.
Two waveforms are distinguishable when $\text{Mismatch}[h_1, h_2]>d/(2\text{SNR}^2)$ is satisfied, 
where $d=13$ is the number of source parameters \cite{Flanagan:1997kp, Lindblom:2008cm}.
This yields a detection threshold of $d/(2\,\text{SNR}^2)= 0.0115$.

The mismatch results for EMRIs evolving with and without DM halos are summarized in Table \ref{table:3}.
These results indicate that, with a fully relativistic treatment of the DM distribution, DM halos with compactness $M/a_0\sim 10^{-5}$ (or a little smaller) can be detectable.
The results also demonstrate the necessity of including the radial pressure.

\begin{table}[htp]
\centering
\caption{Mismatch between EMRI waveforms with and without DM halos over one year observation. 
Left three columns: mismatch of each DM case relative to the vacuum case. Right three columns: pairwise mismatches among the three DM cases.}
\label{Tab:faithfulness} 
\begin{tabular}{|c|c|c|c||c|c|c|}
\hline
$M/a_0$ & case 1 & case 2 & case 3 & case 1/2 & case 1/3 & case 2/3 \\
\hline
$10^{-3}$ & $0.984$ & $0.979$   & $0.957$ & $0.810$ & $0.948$   & $0.928$ \\
$10^{-4}$ & $0.931$ & $0.951$   &  $0.883$ & $0.425$ & $0.860$   &  $0.915$ \\
$10^{-5}$ & $0.843$ & $0.841$   &  $0.638$ & $0.328$ & $0.832$   &  $0.887$\\
\hline          
\end{tabular}
\label{table:3}
\end{table}

\section{Conclusion and Discussion}
\label{conclusion}
By including radial pressure and a fully relativistic treatment of DM energy distribution, 
we constructed a static, spherically symmetric metric describing a Schwarzschild BH embedded in a DM halo. 
We compared three DM prescriptions: case 1 ($\rho_\text{T}$ with radial pressure $p_r$), case 2 ($\rho_\text{J}$ with $p_r=0$), and case 3 ($\rho_\text{C}$ with $p_r=0$). 
Our main findings are

\begin{itemize}
  \item[] Density profiles: $\rho_\text{T}$ and $\rho_\text{J}$ are almost the same; their ratio $\rho_\text{T}/\rho_\text{J}$ increases slightly toward smaller radii, reaching $\approx 1.4$ near the horizon. Both profiles, however, differ significantly from $\rho_\text{C}$. 
For $M/a_0=10^{-4}$ the peak of $\rho_\text{T}$ exceeds that of $\rho_\text{C}$ by approximately six orders of magnitude. 
The peak ratio $p_r/\rho_\text{T}\approx0.01$ for $a_0=100M$ and $M=10^4M_{\rm BH}$.
  \item[] Energy loss: We solved the background geometry for each case and derived orbital-energy loss rates for case 1 including dynamical friction and GW emission. 
GW flux dominates for $r\lesssim14M_{\rm BH}$, 
while dynamical friction dominates for $r\gtrsim16M_{\rm BH}$. 
Overall, halo-induced enhancement of GW radiation remains subdominant to dynamical friction.
  \item[] Orbital dynamics: The presence of DM accelerates the inspiral through dynamical friction and halo-induced modifications to GW emission. 
Radial pressure slightly slows the inspiral relative to the pressureless relativistic case.
Larger compactness $M/a_0$ leads to faster evolution. 
  \item[] Detectability: Both accumulated phase shifts and waveform mismatches demonstrate that relativistic modeling materially changes detectability thresholds.
In particular, DM halos with compactness as small as $M/a_0\sim10^{-5}$
may become observable only when the halo is treated relativistically.
\end{itemize}

Overall, our results show that radial pressure and a fully relativistic description of dark matter can significantly influence EMRI dynamics and gravitational wave observables. 
Accurate modeling of environmental effects is therefore essential for extracting reliable astrophysical and fundamental physics information from future space-borne GW detectors.

\begin{acknowledgments}
This work is supported in part by the National Natural Science Foundation of China  under Grant Nos. 12588101 and 12535002.
\end{acknowledgments}

%

\end{document}